\documentclass[floatfix,noshowpacs,preprintnumbers,twocolumn,amsmath,amssymb,aps,pra,superscriptaddress]{revtex4-2}
\usepackage[T1]{fontenc}
\usepackage{graphicx}
\usepackage{bm}
\usepackage[hidelinks]{hyperref}

\begin{document}
\title{Photon-loss tolerance of the surface-GKP code with anti-photon-loss decoding}
\author{Guan-Ju Yan}
\thanks{These authors contributed equally to this work.}
\affiliation{Center for Integrated Quantum Information Technologies (IQIT), School of Physics and Astronomy and State Key Laboratory of Advanced Optical Communication Systems and Networks, Shanghai Jiao Tong University, Shanghai 200240, China}

\author{Pei-Lin Du}
\thanks{These authors contributed equally to this work.}
\affiliation{TuringQ Co., Ltd., Shanghai 200240, China}

\author{Ke-Ming Hu}
\affiliation{TuringQ Co., Ltd., Shanghai 200240, China}

\author{Jun-Jie He}
\affiliation{TuringQ Co., Ltd., Shanghai 200240, China}

\author{Yu-Ze Zhu}
\affiliation{TuringQ Co., Ltd., Shanghai 200240, China}

\author{Xing-Yu Chen}
\affiliation{Center for Integrated Quantum Information Technologies (IQIT), School of Physics and Astronomy and State Key Laboratory of Advanced Optical Communication Systems and Networks, Shanghai Jiao Tong University, Shanghai 200240, China}

\author{Lin Yang}
\affiliation{TuringQ Co., Ltd., Shanghai 200240, China}

\author{An-Bang Wu}
\email{anbangwu@sjtu.edu.cn}
\affiliation{School of Computer Science, Shanghai Jiao Tong University, Shanghai 200240, China}

\author{Hao Tang}
\email{htang2015@sjtu.edu.cn}
\affiliation{Center for Integrated Quantum Information Technologies (IQIT), School of Physics and Astronomy and State Key Laboratory of Advanced Optical Communication Systems and Networks, Shanghai Jiao Tong University, Shanghai 200240, China}

\author{Xian-Min Jin}
\email{xianmin.jin@sjtu.edu.cn}
\affiliation{Center for Integrated Quantum Information Technologies (IQIT), School of Physics and Astronomy and State Key Laboratory of Advanced Optical Communication Systems and Networks, Shanghai Jiao Tong University, Shanghai 200240, China}
\affiliation{TuringQ Co., Ltd., Shanghai 200240, China}
\affiliation{Chip Hub for Integrated Photonics Xplore (CHIPX), Shanghai Jiao Tong University, Wuxi 214000, China}
\affiliation{Atomology Co., Ltd., Shanghai 200240, China}

\begin{abstract}
Photon loss poses a challenge to fault-tolerant photonic quantum computing. We investigate its effect on finite-energy Gottesman--Kitaev--Preskill (GKP) states by modeling the pure-loss channel directly. We obtain photon-loss threshold estimates for the surface-GKP code using nearest-lattice-point inner GKP decoding and minimum-weight perfect matching for the outer surface code. For GKP states with $\epsilon=0.01$ and $\epsilon=0.05$, corresponding to GKP squeezing of $20$ and $13\,\mathrm{dB}$, the estimated thresholds are $\gamma^\ast=5.5\%$ ($0.24\,\mathrm{dB}$) and $\gamma^\ast=5.9\%$ ($0.26\,\mathrm{dB}$), respectively. We then introduce an anti-photon-loss inner decoder that compensates for the contraction of the measured quadrature. It increases the corresponding threshold estimates to $\gamma^\ast=9.8\%$ ($0.45\,\mathrm{dB}$) and $\gamma^\ast=6.3\%$ ($0.28\,\mathrm{dB}$). Our results highlight the importance of accounting for loss-induced lattice contraction in inner decoding and demonstrate that quadrature compensation enhances the photon-loss tolerance of surface-GKP codes for both high- and moderate-quality GKP states, providing a route toward more photon-loss-robust optical quantum computing.
\end{abstract}
\maketitle

\section{Introduction}
Quantum computing promises speedups for certain problems, but decoherence and noise impede its practical realization~\cite{preskill2018quantum,nielsen2010quantum,ladd2010quantum,divincenzo2000physical}. Among the various noise sources, photon loss affects optical and microwave quantum platforms, including linear photonic quantum computing~\cite{knill2001scheme,kok2007linear,varnava2008good} and bosonic modes in superconducting circuits~\cite{blais2021circuit}. During propagation, storage, and optical processing, loss degrades encoded quantum states.

The Gottesman--Kitaev--Preskill (GKP) code~\cite{gottesman2000encoding,fluhmann2019encoding,campagne2020quantum} offers a way to correct random displacement errors in continuous-variable (CV) systems. Unlike discrete-variable (DV) codes that operate on finite-dimensional physical systems, the GKP code encodes logical information in the phase-space lattice of a bosonic mode, allowing the correction of small shift errors in both position and momentum quadratures~\cite{menicucci2014fault}. Residual shift errors that cross logical decision boundaries give rise to logical Pauli errors, which can be addressed by a higher-level code~\cite{bourassa2021blueprint,noh2020fault,fukui2017analog,stafford2023biased,vuillot2019toric,zhang2021color,raveendran2022finite,zhang2023xzzx}. This makes the GKP code a natural building block for CV fault-tolerant quantum computation (FTQC)~\cite{menicucci2014fault,bourassa2021blueprint,noh2022low,fukui2018tracking,fukui2018high,matsuura2026general,du2025complete}.

The surface code is a leading approach to topological fault tolerance in DV quantum computing~\cite{kitaev2003fault,dennis2002topological}. Its planar layout, local stabilizer measurements, and comparatively high error threshold make it experimentally attractive~\cite{fowler2012surface,wang2011surface,raussendorf2007fault}. Thresholds on the order of $1\%$ have been obtained for standard circuit-level noise models, with the precise value depending on the noise model and decoder~\cite{wang2011surface,stephens2014thresholds}. The surface code is also compatible with nearest-neighbor gates, an important feature for superconducting qubit arrays and other architectures with local connectivity~\cite{barends2014superconducting,kelly2015state}.

The surface-GKP code combines these two codes by replacing each physical qubit of the surface code with a bosonic mode encoded as a GKP qubit~\cite{noh2020fault,noh2022low,fukui2018high,zhang2023xzzx}. In this concatenated architecture, inner GKP correction suppresses shift errors, while the outer surface code corrects residual Pauli errors~\cite{noh2022low,noh2020fault}. However, photon loss is not equivalent to a random-displacement channel~\cite{albert2018performance}. For finite-energy GKP states, the logical channel resulting from loss followed by GKP error correction is not generally a stochastic Pauli channel~\cite{harris2025logical}. Pure loss has a distinctive effect on GKP states: it contracts the phase-space lattice toward the origin and introduces vacuum noise that broadens the initially narrow peaks. A peak farther from the origin undergoes a larger inward shift. These systematic changes in peak position are not captured by simply replacing pure loss with a zero-mean random-displacement channel. The challenge is to determine how a physical loss channel affects the performance of a concatenated surface-GKP code. Although pure loss can be converted into an additive Gaussian random-displacement channel using a suitably chosen quantum-limited, phase-insensitive amplifier~\cite{albert2018performance,noh2019capacity,fukui2021alloptical}, this amplification also proportionally increases the noise already present in the measurement data. We model pure loss directly, retaining both attenuation and vacuum noise without introducing this additional amplification channel~\cite{hastrup2021analysis}.

Here we investigate whether compensating for the loss-induced contraction of GKP quadrature outcomes improves the simulated photon-loss tolerance of the surface-GKP code. Each bosonic mode is subject to a pure-loss channel with loss fraction $\gamma$, described by the beam-splitter transformation $\hat a\to\sqrt{1-\gamma}\,\hat a+\sqrt{\gamma}\,\hat e$, where $\hat e$ denotes an environmental vacuum mode~\cite{weedbrook2012gaussian}. We directly compute the effect of this channel on finite-energy GKP states and determine the resulting inner-decoder error probabilities for the surface-code layer. Using Monte Carlo simulations with minimum-weight perfect matching (MWPM) decoding~\cite{higgott2022pymatching,Higgott2025sparseblossom}, we evaluate the logical failure probability $p_L$ as a function of photon loss and code distance. We further propose an anti-photon-loss (APL) inner decoder that compensates for the inward contraction in the quadrature selected for measurement. We compare the resulting thresholds with those obtained using nearest-lattice-point decoding for $\epsilon=0.01$ and $0.05$. This comparison demonstrates that quadrature compensation enhances the photon-loss tolerance of surface-GKP codes.

\section{Noise model}
\label{sec:noise}
In this section, we describe the two ingredients of the noise model: finite-energy GKP states and photon loss. Finite energy characterizes the prepared states, while the loss channel describes their interaction with the environment during propagation and processing. We treat other operations as ideal apart from the loss assigned in the numerical model. Throughout, we use $\hbar=1$, with
\begin{equation}
 \hat q=\frac{\hat a+\hat a^\dagger}{\sqrt{2}},\qquad
 \hat p=\frac{i(\hat a^\dagger-\hat a)}{\sqrt{2}},\qquad [\hat q,\hat p]=i.
 \label{eq:quadratures}
\end{equation}
The vacuum variance is therefore $1/2$ in each quadrature.

\subsection{Photon loss}
\begin{figure}[t]
 \includegraphics[width=\linewidth]{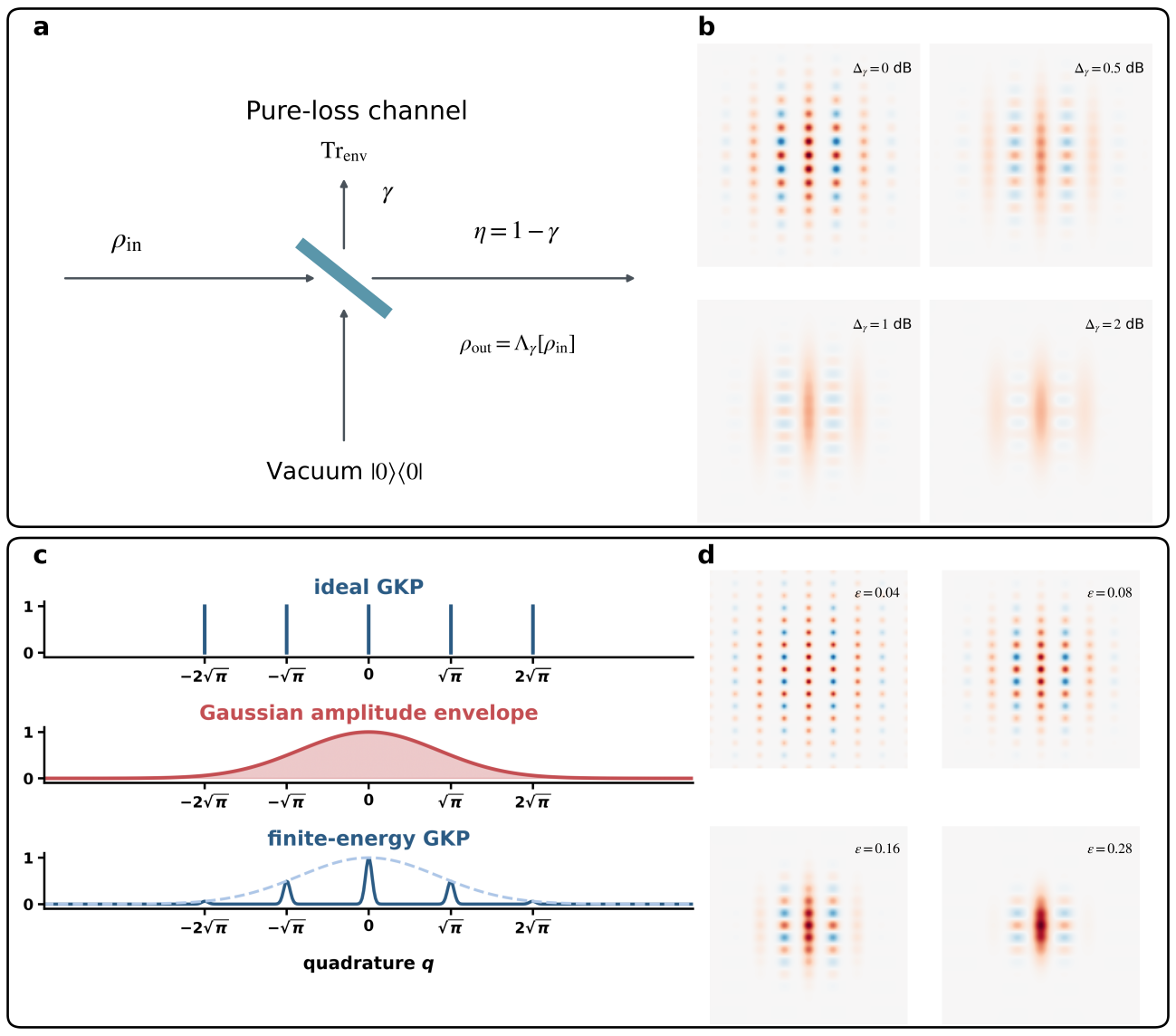}
 \caption{Noise channels and finite-energy effects for GKP states. (a) Beam-splitter model of a pure-loss channel with transmissivity $1-\gamma$, coupling the input to a vacuum environment. (b) Wigner functions at the indicated losses $\Delta_\gamma$, illustrating inward lattice contraction and peak broadening. (c) Schematic $q$-space wavefunction of a GKP $|+\rangle$ state: equal-weight, same-phase ideal peaks at $q=k\sqrt{\pi}$ ($k\in\mathbb Z$). Finite peak widths, the Gaussian amplitude envelope, and vertical scales are illustrative. (d) Wigner functions for the indicated envelope parameters $\epsilon$, illustrating changes in peak width and envelope extent.}
 \label{fig:2}
\end{figure}

We consider a single-mode pure-loss channel with loss fraction $\gamma\in[0,1]$ and transmissivity $\eta=1-\gamma$. A standard physical model is a beam splitter coupling the system to an environmental vacuum mode, as shown in Fig.~\ref{fig:2}(a)~\cite{weedbrook2012gaussian}. The channel acts on an input state $\rho$ as~\cite{kim1995phase}
\begin{equation}
 \Lambda_\gamma[\rho]=\operatorname{Tr}_{\mathrm{env}}\!\left[
 U_{\mathrm{BS},\gamma}(\rho\otimes|0\rangle\langle0|)
 U_{\mathrm{BS},\gamma}^{\dagger}\right],
 \label{eq:loss_channel_beam_splitter}
\end{equation}
where we choose the beam-splitter phase convention
\begin{equation}
 U_{\mathrm{BS},\gamma}
 =e^{\theta_\gamma(\hat a^\dagger\hat a_{\mathrm{env}}-\hat a_{\mathrm{env}}^\dagger\hat a)},
 \qquad \theta_\gamma=\arccos\sqrt{1-\gamma}.
\end{equation}
The corresponding transformations of the amplitude and phase quadratures are
\begin{align}
 \hat q_{\mathrm{out}}&=\sqrt{1-\gamma}\,\hat q_{\mathrm{in}}
       +\sqrt{\gamma}\,\hat q_{\mathrm{env}},\label{eq:loss_q}\\
 \hat p_{\mathrm{out}}&=\sqrt{1-\gamma}\,\hat p_{\mathrm{in}}
       +\sqrt{\gamma}\,\hat p_{\mathrm{env}}.\label{eq:loss_p}
\end{align}
These equations display the two physical effects of loss explicitly: both quadratures are attenuated by $\sqrt{1-\gamma}$, and vacuum fluctuations enter through the environmental mode. Loss is expressed in decibels as
\begin{equation}
 \Delta_\gamma=-10\log_{10}(1-\gamma)\quad\text{(dB)}.
 \label{eq:loss_db}
\end{equation}

The same effects can be separated in the Wigner representation. We first define the attenuated distribution
\begin{equation}
 W_{\mathrm{att}}(q,p)=\frac{1}{1-\gamma}
 W_{\mathrm{in}}\!\left(\frac{q}{\sqrt{1-\gamma}},
                       \frac{p}{\sqrt{1-\gamma}}\right).
 \label{eq:loss_contraction}
\end{equation}
This coordinate scaling maps a reference peak center $(q_k,p_k)$ to $(\sqrt{1-\gamma}\,q_k,\sqrt{1-\gamma}\,p_k)$, making the inward contraction explicit. The factor $1/(1-\gamma)$ preserves normalization as the phase-space area contracts. The physical output is obtained by including the environmental vacuum noise:
\begin{align}
 W_{\mathrm{out}}(q,p)&=\int dq'\,dp'\,
 G_\gamma(q-q',p-p')W_{\mathrm{att}}(q',p'),
 \label{eq:loss_wigner}\\
 G_\gamma(q,p)&=\frac{1}{\pi\gamma}
 e^{-(q^2+p^2)/\gamma}.\label{eq:loss_kernel}
\end{align}
Equations~\eqref{eq:loss_contraction}--\eqref{eq:loss_kernel} describe contraction followed by convolution and apply for $0<\gamma<1$, with the endpoints defined by their limits. The attenuation step is a useful mathematical decomposition; the convolution is essential for the full pure-loss channel.

For an initially narrow Gaussian peak centered at $q_k$ with variance $v_q$, these transformations give
\begin{equation}
 q_k\longrightarrow\sqrt{1-\gamma}\,q_k,
 \qquad v_q\longrightarrow(1-\gamma)v_q+\frac{\gamma}{2},
 \label{eq:peak_loss}
\end{equation}
with analogous transformations for $p$. Since the narrow GKP peaks have subvacuum variance, the added vacuum term broadens them. Meanwhile, their displacement from the original positions is directed toward the origin and has magnitude $(1-\sqrt{1-\gamma})|q_k|$, as shown in Fig.~\ref{fig:2}(b). Both effects can increase the probability of crossing a logical decision boundary.

As $\gamma\to1$, a finite-energy input approaches the vacuum state,
\begin{equation}
 \lim_{\gamma\to1}W_{\mathrm{out}}(q,p)
 =W_{\mathrm{vac}}(q,p)=\frac{1}{\pi}e^{-(q^2+p^2)}.
\end{equation}
This limit reflects the loss of the input photons to the environment. Detector inefficiency can also be represented by mixing the measured mode with vacuum before ideal homodyne detection. For a propagation loss $\gamma_0$ followed directly by a detector of efficiency $\eta_{\mathrm{det}}$, the total loss is $\gamma_{\mathrm{tot}}=1-\eta_{\mathrm{det}}(1-\gamma_0)$.

\subsection{Finite-energy GKP states}
To describe finite-energy GKP states, we use the envelope operator~\cite{bourassa2021fast,hastrup2021analysis}
\begin{equation}
 \hat E(\epsilon)=e^{-\epsilon\hat n},\qquad
 \hat n=\hat a^\dagger\hat a,
 \label{eq:envelope}
\end{equation}
where $\epsilon>0$ controls the departure from an ideal GKP state. Applying this operator to an ideal computational-basis state gives
\begin{equation}
 |\mu^\epsilon\rangle_{\mathrm{GKP}}
 =\frac{e^{-\epsilon\hat n}|\mu\rangle_{\mathrm{GKP}}}
 {\sqrt{_{\mathrm{GKP}}\langle\mu|e^{-2\epsilon\hat n}|\mu\rangle_{\mathrm{GKP}}}},
 \qquad\mu=0,1.
 \label{eq:finite_gkp}
\end{equation}
The operator suppresses high-photon-number components, producing finite-width peaks and a decaying envelope. Figure~\ref{fig:2}(c) schematically illustrates the $q$-space wavefunction of a logical GKP $|+\rangle$ state, regularized as in Eq.~\eqref{eq:finite_gkp}, not by envelope multiplication alone. Decreasing $\epsilon$ narrows the peaks and extends the envelope. The state approaches the ideal limit as $\epsilon\to0$.

For the damping convention in Eq.~\eqref{eq:envelope}, the nominal peak variance is $v_{\mathrm{peak}}=\tfrac12\tanh\epsilon$. Relative to the vacuum variance $v_{\mathrm{vac}}=1/2$, the corresponding GKP squeezing is~\cite{hastrup2021analysis}
\begin{equation}
 s_{\mathrm{GKP}}=-10\log_{10}(\tanh\epsilon)\quad\text{(dB)}.
 \label{eq:squeezing}
\end{equation}
As shown in Fig.~\ref{fig:2}(d), increasing $\epsilon$ reduces the number of appreciable peaks and broadens them. The envelope operator defines a finite-energy input state rather than a dynamical loss channel. Unlike pure loss, it does not arise from tracing out a vacuum environment. Its exact action also produces a small shift of the peak centers, so finite-energy broadening and physical attenuation should be distinguished even though both modify the phase-space pattern. In the following, we retain the finite-energy structure when evaluating the action of photon loss.

\section{The surface-GKP code}
\label{sec:surface}
In this section, we introduce the surface-GKP code, which concatenates the single-mode GKP code with the surface code~\cite{fowler2012surface,noh2020fault}. We first specify the GKP lattice, then describe the rotated surface-code layout and the two stages of decoding.

The GKP code encodes a logical qubit into a single bosonic mode using a phase-space lattice~\cite{gottesman2000encoding}. With the convention $[\hat q,\hat p]=i$ used throughout this work, the computational-basis states are
\begin{equation}
 |\mu\rangle_{\mathrm{GKP}}\propto
 \sum_{n\in\mathbb Z}|q=(2n+\mu)\sqrt{\pi}\rangle,
 \qquad\mu=0,1.
 \label{eq:gkp_general}
\end{equation}
Adjacent logical grid points are separated by $\ell=\sqrt{\pi}$, while peaks within either computational-basis state are separated by $2\ell$. The corresponding nearest-grid decision boundaries lie halfway between adjacent logical grid points, at $(k+1/2)\ell$ for $k\in\mathbb Z$.

The surface code is a topological stabilizer code defined on a two-dimensional lattice~\cite{kitaev2003fault,dennis2002topological}. We use the rotated surface code, which requires $d^2$ data qubits and $d^2-1$ syndrome qubits for code distance $d$. In the absence of syndrome-measurement errors, the code can correct any Pauli error of weight at most $\lfloor(d-1)/2\rfloor$. Its stabilizers are of two types, $Z$ and $X$, each measured using a dedicated syndrome qubit~\cite{raussendorf2007fault}.

\begin{figure}[t]
 \includegraphics[width=\linewidth]{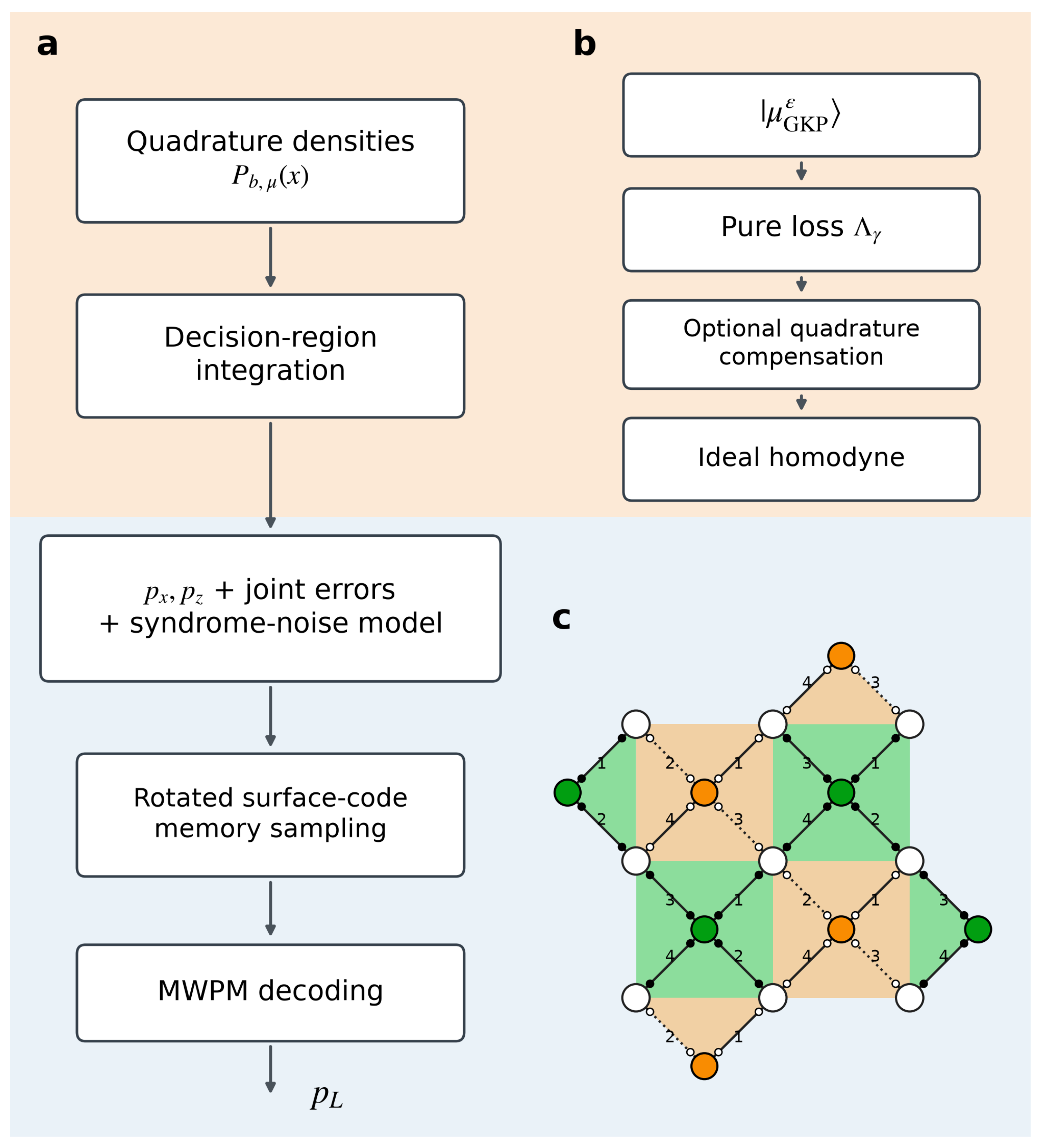}
 \caption{Effective surface-GKP code. (a) Numerical workflow: conditional quadrature error probabilities are used to construct a qubit-noise model for a rotated surface code; sampled detection events are then decoded by MWPM to obtain $p_L$. The reduction requires a joint Pauli-error assignment and a syndrome-noise model in addition to the two marginal error probabilities. (b) Bosonic operations explicitly represented in the readout model: finite-energy preparation, one pure-loss channel, optional quadrature compensation, and ideal homodyne detection. The gain is $1/\sqrt{1-\gamma}$ for the measured quadrature; it can be implemented by classical rescaling. (c) Conceptual distance-three code layout. White circles represent data modes, and colored circles represent syndrome modes of the two stabilizer types; the numbers indicate the proposed interaction order.}
 \label{fig:1}
\end{figure}

The concatenation uses the GKP code to address continuous quadrature errors and the surface code to correct the remaining logical Pauli errors. As shown in Fig.~\ref{fig:1}(c), the architecture contains $d^2$ data GKP qubits and $(d^2-1)/2$ syndrome GKP qubits for each stabilizer type. The total is $2d^2-1$ GKP states, excluding additional resources for inner GKP correction and state preparation.

Figure~\ref{fig:1}(a) separates the single-mode calculation from the effective qubit-level simulation. The original numerical workflow uses DeepQuantum~\cite{he2025deepquantum} for quadrature marginals, interval integration for decision probabilities, Stim~\cite{gidney2021stim} for the qubit-level circuit, and PyMatching~\cite{higgott2022pymatching,Higgott2025sparseblossom} for MWPM.

For a readout basis $b=q,p$, let $P_{b,\mu}(x)$ be the normalized probability density conditional on the prepared binary label $\mu$. The $q$ tests use the damped logical $|0\rangle$ and $|1\rangle$ states; the corresponding $p$ tests use the normalized damped $|+\rangle$ and $|-\rangle$ states. With $\ell=\sqrt{\pi}$ and
\begin{equation}
 \mathcal D_j=\bigcup_{k\in\mathbb Z}
 [(2k+j-\tfrac12)\ell,(2k+j+\tfrac12)\ell),
 \label{eq:decision_sets}
\end{equation}
the conditional readout error probabilities and their equal-prior averages are
\begin{align}
 e_{b,\mu}&=\int_{\mathcal D_{1-\mu}}P_{b,\mu}(x)\,dx,
 \label{eq:conditional_errors}\\
 \bar p_x&=\tfrac12(e_{q,0}+e_{q,1}),\qquad
 \bar p_z=\tfrac12(e_{p,0}+e_{p,1}).
 \label{eq:basis_average}
\end{align}
Conditional and averaged error probabilities must be distinguished when passing a numerical probability to the outer code. The notation $p_x,p_z$ below denotes the chosen bit- and phase-error marginals; equal-prior averaging would set them to $\bar p_x,\bar p_z$.

Even after this choice, the two marginals do not uniquely determine a Pauli channel. A stochastic Pauli surrogate has the form
\begin{equation}
 \mathcal P(\rho)=P_I\rho+P_XX\rho X+P_YY\rho Y+P_ZZ\rho Z,
 \label{eq:pauli_surrogate}
\end{equation}
where $p_x=P_X+P_Y$ and $p_z=P_Z+P_Y$, respectively. Writing $c=P_Y$ gives
\begin{align}
 &P_I=1-p_x-p_z+c,P_X=p_x-c,\nonumber\\
 &P_Z=p_z-c,P_Y=c,
 \label{eq:joint_pauli}\\
 &\max(0,p_x+p_z-1)\le c\le\min(p_x,p_z).\nonumber
\end{align}
Independent bit and phase flips correspond to the additional assumption $c=p_xp_z$; setting $P_Y=0$ is a different model and is possible only when $p_x+p_z\le1$. Separate syndrome-outcome error probabilities $m_X,m_Z$ are also needed whenever syndrome measurements are noisy. They are not determined by the data-mode marginals. Spatial and temporal correlations, ancilla noise, and the locations where these probabilities are applied likewise belong to the outer-circuit specification.

The numerical reduction therefore follows four stages:
\begin{enumerate}
 \item Prepare a single-mode finite-energy test state, apply pure loss once, and calculate its quadrature density.
 \item Integrate the density over incorrect decision regions, using either nearest-lattice-point or APL decoding.
 \item Use the specified effective Pauli-error and syndrome-noise models in the rotated qubit-level circuit, sample detection events, and decode them using MWPM.
 \item Compare the inferred recovery with the sampled logical observable to estimate the failure probability of the surface-GKP code.
\end{enumerate}
Steane-type ancilla coupling is part of a possible physical realization of the concatenated architecture; it is not an operation simulated by this single-mode-to-qubit reduction. Consequently, the effective-channel comparison does not establish the residual logical channel of a full bosonic correction cycle. In particular, Eqs.~\eqref{eq:conditional_errors}--\eqref{eq:joint_pauli} specify readout statistics and possible Pauli reductions, but do not derive the latter from a trace-preserving bosonic recovery.

\section{Anti-photon-loss inner decoder}
\label{sec:apl}
\begin{figure}[t]
 \includegraphics[width=\linewidth]{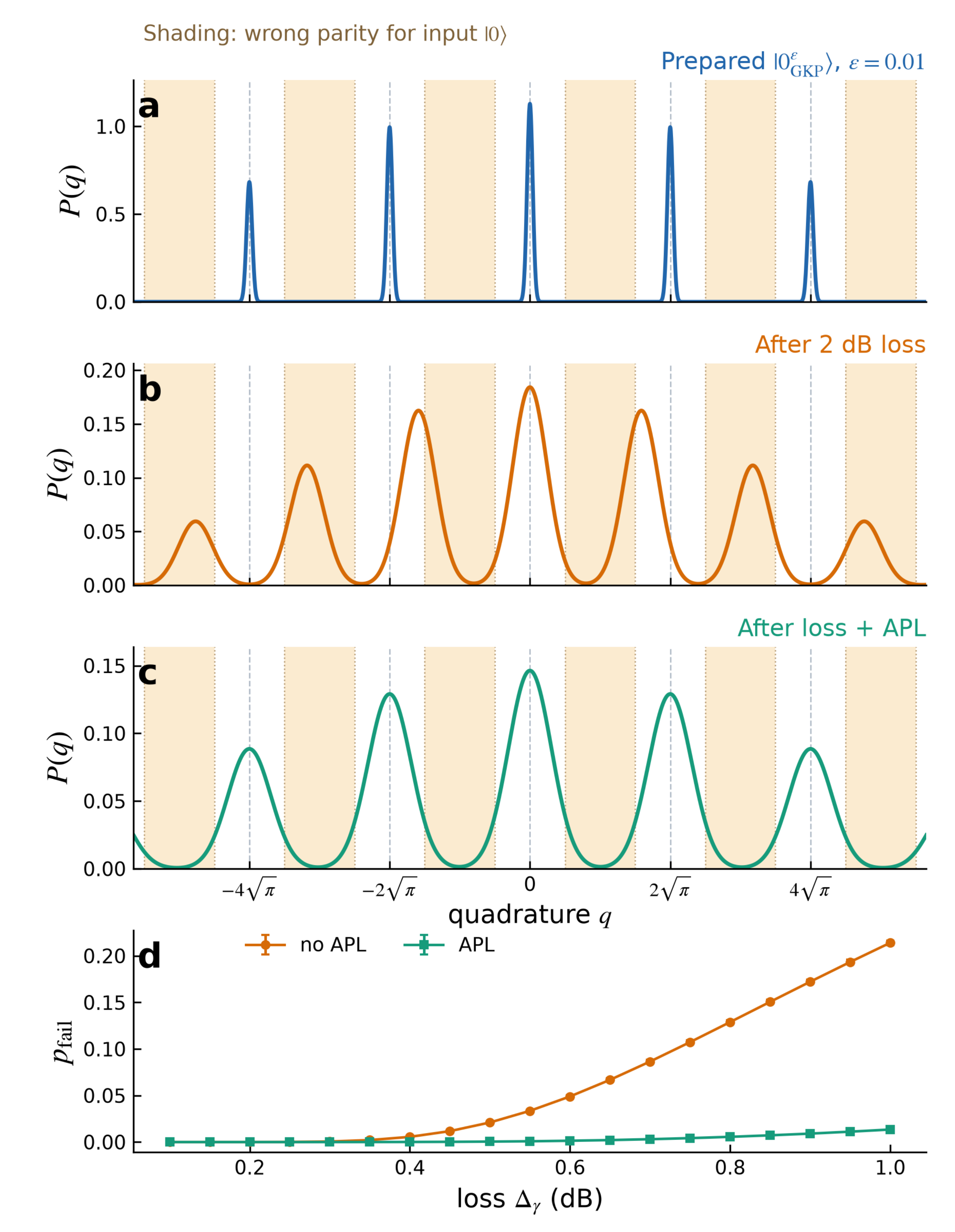}
 \caption{APL inner decoder. (a)--(c) $q$-quadrature distributions for a finite-energy GKP state with $\epsilon=0.01$: (a) the prepared state; (b) after $2\,\mathrm{dB}$ of photon loss; and (c) after the same loss followed by $\hat{S}(r_{\mathrm{comp}})$. This loss value illustrates the deformation and does not represent a threshold operating point. Dashed lines mark reference grid positions $q=k\ell$, not the decision boundaries $q=(k+1/2)\ell$, where $\ell=\sqrt{\pi}$. (d) Inner-decoder failure probability $p_{\mathrm{fail}}$ versus photon loss $\Delta_\gamma$, with APL (green squares) and without APL (orange circles). APL reduces the failure probability, with a more pronounced reduction at higher losses within the plotted range. Compensation restores the measured peak-position scale but does not remove loss-induced vacuum noise.}
 \label{fig:3}
\end{figure}

A key observation for combating photon loss is that it contracts the GKP lattice toward the origin while introducing vacuum noise. Figure~\ref{fig:3}(a) shows the $q$ distribution of a prepared GKP state with $\epsilon=0.01$; Fig.~\ref{fig:3}(b) shows the inward shift and broadening after loss. Equation~\eqref{eq:loss_contraction} identifies the contraction factor as $\sqrt{1-\gamma}$. The resulting mismatch between peak positions and the fixed decoding grid can increase the residual error probability after homodyne decoding.

For nearest-lattice-point decoding, a quadrature outcome is assigned to the nearest multiple of $\ell$. An outcome in the interval $[(k-1/2)\ell,(k+1/2)\ell)$ is assigned the parity of $k$. Loss moves outer peaks toward these decision boundaries, even when the peaks of the prepared state are narrow. Compensating for their inward motion therefore addresses a source of decoding errors that is distinct from the finite width of the peaks.

To address this effect, we propose an \emph{anti-photon-loss (APL) inner decoder}. The core idea is to apply a quadrature-dependent counter-squeezing operation before homodyne readout. Related phase-sensitive amplification and measurement-outcome rescaling have been studied for loss compensation in GKP-based quantum communication~\cite{fukui2021alloptical,Hillmann}. Here we evaluate quadrature compensation in a surface-GKP code. We define
\begin{equation}
 \hat S(r)=\exp\!\left[\frac r2(\hat a^{\dagger2}-\hat a^2)\right],
 \label{eq:squeeze_gate}
\end{equation}
which transforms the quadratures as
\begin{equation}
 \hat S^\dagger(r)\hat q\hat S(r)=e^r\hat q,
 \qquad
 \hat S^\dagger(r)\hat p\hat S(r)=e^{-r}\hat p.
 \label{eq:squeeze_quadratures}
\end{equation}
For $q$ readout, the loss-compensating value is
\begin{equation}
 r_{\mathrm{comp}}=-\frac12\ln(1-\gamma),
 \label{eq:compensation}
\end{equation}
so that $e^{r_{\mathrm{comp}}}\sqrt{1-\gamma}=1$. Figure~\ref{fig:3}(c) shows the compensated $q$ distribution. For $p$ readout, the squeezing orientation is reversed by using $\hat S(-r_{\mathrm{comp}})$. In each case, compensation restores the pre-loss scale of the quadrature selected for measurement.

The action of APL can be seen directly by combining Eqs.~\eqref{eq:loss_q} and~\eqref{eq:compensation}:
\begin{equation}
 \hat q_{\mathrm{APL}}=\hat q_{\mathrm{in}}
       +\sqrt{\frac{\gamma}{1-\gamma}}\,\hat q_{\mathrm{env}}.
 \label{eq:apl_q}
\end{equation}
The deterministic attenuation is canceled, while vacuum noise remains. Thus, APL realigns the measured peak pattern with the decoding grid rather than reversing the loss channel. A peak with pre-loss variance $v_q$ has variance $v_q+\gamma/[2(1-\gamma)]$ after compensation. Whether this realignment improves decoding depends on the balance between the benefit of realignment and the amplified vacuum noise.

The corresponding probability distribution for the compensated outcome $y$ is
\begin{equation}
 P_{\mathrm{APL}}(y)=\sqrt{1-\gamma}\,
 P_{\mathrm{loss}}\!\left(\sqrt{1-\gamma}\,y\right).
 \label{eq:apl_probability}
\end{equation}
This expression gives a direct way to evaluate the inner-decoder error probability. For ideal homodyne readout, it also shows that the same decision statistics can be obtained by rescaling the classical outcome or, equivalently, by contracting the decision intervals~\cite{fukui2021alloptical}. Thus, the ideal-homodyne calculations assess a change in the decision scale; they do not distinguish a physical squeezing operation from classical rescaling of the same outcomes.

The choice of squeezing orientation is essential: each readout uses the operation appropriate to the quadrature being measured. The conjugate quadrature undergoes the opposite scaling, as required by Eq. (22). This quadrature-selective operation is the basis of APL. Figure 3(d) compares the inner-decoder failure probability $p_{\mathrm{fail}}$ with and without APL as a function of photon loss. APL reduces the failure probability, with a more pronounced reduction at higher losses within the plotted range. At the largest estimated surface-code threshold discussed below, $\gamma=9.8\%$, the compensating parameter is only $r_{\mathrm{comp}}\simeq0.052$, corresponding to about $0.45\,\mathrm{dB}$ of quadrature anti-squeezing. In the ideal-homodyne model considered here, this corresponds to a modest rescaling of the measured quadrature.

\section{Main results}
\label{sec:results}
\subsection{Numerical simulation}
\begin{figure*}[t]
 \includegraphics[width=0.91\textwidth]{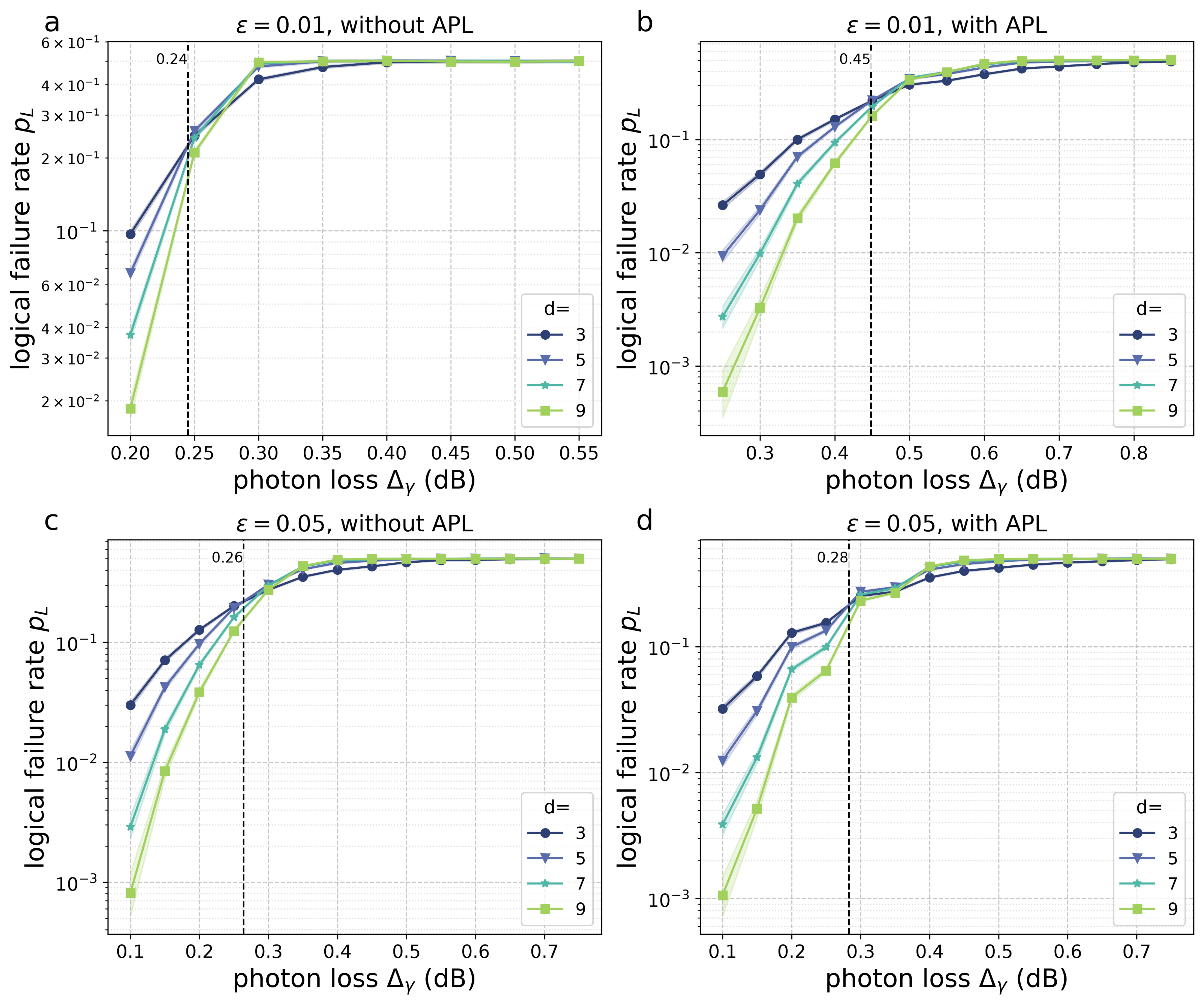}
 \caption{Logical failure probability $p_L$ of the surface-GKP code versus photon loss for code distances $d=3,5,7,9$. Each simulation trial comprises $d+1$ noisy stabilizer-measurement rounds and a final ideal round. (a) High-quality GKP states, $\epsilon=0.01$, without APL. (b) The same states with APL. (c) Moderate-quality GKP states, $\epsilon=0.05$, without APL. (d) The same states with APL. The respective finite-distance threshold estimates are $\gamma^*=5.5\%$ ($0.24\,\mathrm{dB}$), $9.8\%$ ($0.45\,\mathrm{dB}$), $5.9\%$ ($0.26\,\mathrm{dB}$), and $6.3\%$ ($0.28\,\mathrm{dB}$). The results are obtained from $100{,}000$ Monte Carlo trials. Shaded regions show the corresponding error ranges of $p_L$; they do not specify threshold uncertainty. Dashed lines mark the crossover estimates in decibels.}
 \label{fig:Fig4}
\end{figure*}

We estimate the photon-loss threshold of the surface-GKP code numerically. The simulation combines the state-level treatment of photon loss with Monte Carlo sampling of the outer surface code, using up to 100,000 trials per parameter setting, as illustrated in Fig.~\ref{fig:1}(a). For each pair $(\epsilon,\gamma)$, we calculate the quadrature distributions with and without APL and integrate them over the inner-decoder decision regions to obtain the effective bit- and phase-error probabilities.

For example, if $P_0(q)$ is the readout distribution for a prepared logical $|0\rangle_{\mathrm{GKP}}$ state, its wrong-parity probability is
\begin{equation}
 p_{\mathrm{fail}}^{(0)}=
 \sum_{k\in\mathbb Z}\int_{(2k+1/2)\ell}^{(2k+3/2)\ell}
 P_0(q)\,dq.
 \label{eq:failure_probability}
\end{equation}
The integration intervals are the odd-parity decision regions for this logical input. Thus, the decoding error is determined by the total probability mass in the incorrect regions, not by peak positions alone. Using the APL distribution of Eq.~\eqref{eq:apl_probability} gives the corresponding compensated probability. These integrated probabilities provide the noise parameters used in the outer simulation, connecting the quadrature distributions illustrated in Fig.~\ref{fig:3}(a)--(c) to the logical performance of the surface code.

Following the repeated-syndrome approach of Ref.~\cite{noh2020fault}, we perform $d+1$ rounds of noisy stabilizer measurements followed by a final ideal measurement round. The resulting syndrome history is decoded with MWPM~\cite{higgott2022pymatching,Higgott2025sparseblossom}. Separate matching problems are constructed for the two stabilizer types, including the temporal information from repeated measurements. The calculated error probabilities determine the matching weights. We evaluate the logical failure probability $p_L$ for code distances $d=3,5,7,9$. Here, $p_L$ is the failure probability of the surface-GKP code over the full simulated measurement sequence, rather than a per-round error probability. We estimate photon-loss thresholds from the crossover of the finite-distance curves, below which increasing $d$ suppresses logical failures. 

\subsection{High- and moderate-quality GKP states under photon loss}

Figure~\ref{fig:Fig4} compares nearest-lattice-point and APL decoding at each preparation quality. The logical failure probabilities are obtained from $100{,}000$ Monte Carlo trials. The shaded regions indicate the corresponding error ranges of $p_L$, rather than uncertainty intervals for the threshold locations.

For higher-quality GKP states, the prepared peaks are narrower, but attenuation still shifts their centers away from the fixed decoding grid. Figure~\ref{fig:Fig4}(a) gives
\begin{equation}
 \gamma^*\simeq5.5\%,\qquad \Delta_\gamma^*\simeq0.24\,\mathrm{dB}.
\end{equation}
The lower estimated uncompensated threshold relative to the $\epsilon=0.05$ case suggests that peak positions matter alongside peak widths. A smaller $\epsilon$ narrows the peaks while extending the envelope, so more distant peaks can experience larger inward shifts. This competition is a possible explanation for why narrower prepared peaks do not yield a higher threshold with nearest-lattice-point decoding in these simulations.

For moderate-quality GKP states, both finite-energy broadening and photon loss contribute to the decoding error. Without APL, Fig.~\ref{fig:Fig4}(c) gives a threshold of
\begin{equation}
 \gamma^*\simeq5.9\%,\qquad \Delta_\gamma^*\simeq0.26\,\mathrm{dB}.
\end{equation}
Below this estimated crossover, the plotted logical failure probabilities decrease as the code distance increases.

Applying APL to the high-quality states compensates for the loss-induced contraction in the measured quadrature. As shown in Fig.~\ref{fig:Fig4}(b), the threshold increases to
\begin{equation}
 \gamma^*\simeq9.8\%,\qquad \Delta_\gamma^*\simeq0.45\,\mathrm{dB}.
\end{equation}
The point estimates of $\gamma^*$ differ by $4.3$ percentage points, corresponding to a relative increase of approximately $79\%$ over the uncompensated estimate.

APL also improves the loss tolerance of moderate-quality states. Figure~\ref{fig:Fig4}(d) gives
\begin{equation}
 \gamma^*\simeq6.3\%,\qquad \Delta_\gamma^*\simeq0.28\,\mathrm{dB},
\end{equation}
corresponding to an increase in $\gamma^*$ of $0.4$ percentage points, or a relative increase of approximately $7\%$ over the estimate without APL. The plotted finite-distance crossovers shift to higher loss for both values of $\epsilon$. The shift is larger for $\epsilon=0.01$ than for $\epsilon=0.05$ in these simulations. Table~\ref{tab:thresholds} summarizes the four finite-distance crossover estimates.

\begin{table}[t]
 \caption{Photon-loss threshold estimates with and without APL decoding. Decibel values are obtained from Eq.~\eqref{eq:loss_db}.}
 \label{tab:thresholds}
 \begin{ruledtabular}
 \begin{tabular}{cccc}
 $\epsilon$ & Inner decoder & $\gamma^*$ & $\Delta_\gamma^*$ (dB)\\
 \hline
 0.01 & Without APL & $5.5\%$ & 0.24\\
 0.01 & APL & $9.8\%$ & 0.45\\
 0.05 & Without APL & $5.9\%$ & 0.26\\
 0.05 & APL & $6.3\%$ & 0.28
 \end{tabular}
 \end{ruledtabular}
\end{table}

\subsection{Loss-dependent squeezing threshold with APL decoding}
\begin{figure}[t] 
\includegraphics[width=\linewidth]{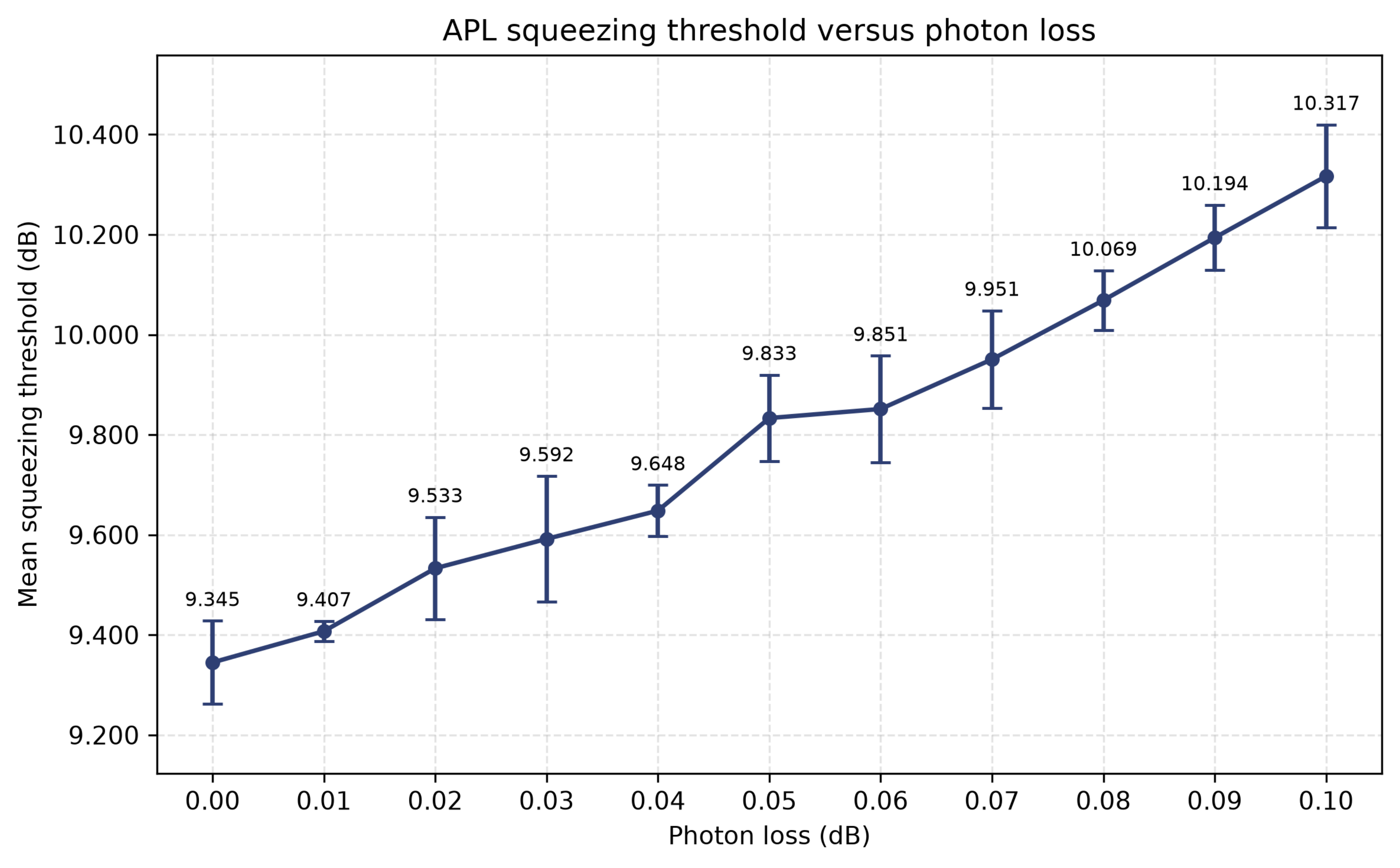} 
\caption{Estimated GKP squeezing threshold with APL decoding versus photon loss. The horizontal axis is the channel attenuation $\Delta_\gamma$ of Eq.~\eqref{eq:loss_db}; the vertical axis is the threshold preparation squeezing $s_{\mathrm{GKP}}^*$, using the convention of Eq.~\eqref{eq:squeezing}. Markers and error bars represent the mean and one standard deviation, respectively. Labels give the mean values, which increase from $9.345\,\mathrm{dB}$ at zero loss to $10.317\,\mathrm{dB}$ at $0.10\,\mathrm{dB}$ loss. The line connects the sampled points as a guide to the eye.} 
\label{fig:5}
\end{figure}

Figure~\ref{fig:5} examines how photon loss changes the GKP preparation quality required by the surface-GKP code with APL decoding. In contrast to Fig.~\ref{fig:Fig4}, which varies loss at fixed preparation quality, this figure shows the estimated squeezing threshold $s_{\mathrm{GKP}}^*(\Delta_\gamma)$ at each fixed loss. Here, the squeezing threshold characterizes the crossover to the regime in which increasing the code distance suppresses the logical failure probability. The horizontal axis is the channel attenuation $\Delta_\gamma=-10\log_{10}(1-\gamma)$, whereas the vertical axis is the preparation squeezing defined in Eq.~\eqref{eq:squeezing}; these are distinct quantities despite both being expressed in decibels.

The estimated mean threshold increases from $9.345\,\mathrm{dB}$ at zero loss to $9.833\,\mathrm{dB}$ at $\Delta_\gamma=0.05\,\mathrm{dB}$ and $10.317\,\mathrm{dB}$ at $\Delta_\gamma=0.10\,\mathrm{dB}$. The largest plotted attenuation corresponds to a loss fraction $\gamma\simeq2.28\%$. The markers and error bars represent the means and one standard deviation, respectively, using up to 1,000,000 simulation samples. The overlapping error bars do not establish a resolved difference between every pair of adjacent loss settings.
The upward trend is consistent with the residual vacuum noise after APL. Equation~\eqref{eq:apl_q} shows that rescaling cancels the deterministic contraction of the measured quadrature but leaves an added variance $\gamma/[2(1-\gamma)]$. This contribution increases with loss, so narrower prepared peaks, and hence greater GKP squeezing, are needed to maintain sufficiently small residual errors for the outer code. APL therefore compensates for the lattice mismatch without removing the preparation-quality cost of loss. At $\gamma=0$, the compensation gain is unity and APL coincides with fixed-grid decoding; the zero-loss point provides the finite-energy reference for the curve.

\section{Discussion and Conclusions}
The comparison in Fig.~\ref{fig:Fig4} indicates that compensating for the contraction of the measured GKP quadrature can improve the loss tolerance of the simulated surface-GKP code. In the absence of compensation, the estimated threshold for the narrower-peaked $\epsilon=0.01$ state is slightly lower than that for $\epsilon=0.05$. This is because the wider envelope at smaller $\epsilon$ gives more weight to distant peaks, which undergo larger inward displacements under loss. This behavior is consistent with the peak motion described by Eq.~\eqref{eq:peak_loss}. Together, Figs.~\ref{fig:Fig4} and~\ref{fig:5} describe complementary aspects of the loss--preparation tradeoff: the former estimates the tolerable loss at selected preparation qualities, while the latter estimates the required squeezing as loss is varied.

The threshold estimates depend on the prepared-state convention, the assignment of pure loss, and the effective Pauli channels used in the surface-code simulation. Logical-channel analyses of approximate GKP error correction distinguish finite-energy damping from stochastic displacement-noise descriptions~\cite{jafarzadeh2025logical}. The effective Pauli channels used here retain quadrature decision probabilities rather than the full logical channel, which can be non-Pauli under pure loss~\cite{harris2025logical}. Whereas Hillmann et al. \cite{Hillmann} evaluated GKP readout with loss-rescaled decision boundaries in teleportation-based error-correction circuits using average gate fidelity, we quantify the effect of quadrature compensation on photon-loss threshold estimates in an effective surface-GKP model with repeated syndrome measurements and minimum-weight perfect-matching decoding.

In conclusion, we modeled pure loss on finite-energy GKP states and used the resulting inner-decoder error probabilities in a surface-code simulation. The estimated loss thresholds without APL are $5.5\%$ for $\epsilon=0.01$ and $5.9\%$ for $\epsilon=0.05$. With quadrature-dependent compensation, the corresponding estimates are $9.8\%$ and $6.3\%$. Within the stated model, accounting for lattice contraction improves the photon-loss tolerance for both state qualities. Our work identifies a directly testable decoding choice for future models with realistic operations.

\begin{acknowledgments}
This research is supported by the National Key R$\&$D Program of China (Grant Nos. 2024YFB4504005 and 2024YFA1409300); the National Natural Science Foundation of China (NSFC) (Grant Nos. 62235012, 12304342, 12574549, 12574542, 125B1033, T2600595, and 62605257); the Quantum Science and Technology-National Science and Technology Major Project (Grant Nos. 2021ZD0301500 and 2021ZD0300700); the Science and Technology Commission of Shanghai Municipality (STCSM) (Grant Nos. 2019SHZDZX01, 24ZR1438700, 24ZR1430700, and 24LZ1401500); the Shanghai Municipal Commission of Economy and Informatization (No. 2025-GZL-RGZN-02047); and the Frontier Technologies R$\&$D Program of Jiangsu (Grant No. SBF20250000094). X.-M. J. acknowledges additional support from a Shanghai talent program and support from the Zhiyuan Innovative Research Center of Shanghai Jiao Tong University. H. T. acknowledges an open research fund from the State Key Laboratory of Photonics and Communications and additional support from the Yangyang Development Fund. A.-B. W. was also supported by the National Natural Science Foundation of China (Grant No. 62502300).
\end{acknowledgments}
\bibliographystyle{apsrev4-2}
\bibliography{ref}
\end{document}